\documentclass[journal,comsoc,10pt]{IEEEtran}

\usepackage[T1]{fontenc}

\usepackage{makecell}
\usepackage{cite}
\usepackage[colorlinks,
linkcolor=red,
anchorcolor=blue,
citecolor=green
]{hyperref}

\ifCLASSINFOpdf
\else
\fi
\usepackage{amsmath}
\usepackage{diagbox}

\usepackage{ amssymb }
\usepackage{amsfonts}
\usepackage{caption}
\usepackage{graphicx}
\usepackage{lettrine}
\usepackage{epstopdf}
\usepackage{textcomp,booktabs}
\usepackage[usenames,dvipsnames]{color}
\usepackage{colortbl}
\definecolor{mygray}{gray}{.9}
\definecolor{mypink}{rgb}{.99,.91,.95}
\definecolor{mycyan}{cmyk}{.3,0,0,0}
\usepackage{pifont}
\usepackage{subfigure}
\usepackage{multirow}
\usepackage{url}
\usepackage{color}
\usepackage{threeparttable}
\usepackage{setspace}
\usepackage{lineno}
\usepackage{tabularx}
\usepackage[dvipsnames,usenames]{color}

\usepackage{bm}

\usepackage{algorithm} 
\usepackage{algorithmic} 

\usepackage{dblfloatfix}

\usepackage[dvipsnames,usenames]{color}
\usepackage[usenames,dvipsnames]{color}

\definecolor{light-gray}{gray}{0.90}
\begin{document}


	\title{Position-Aided Semantic Communication for Efficient Image Transmission:  Design, Implementation, and Experimental Results}

	\author{Peiwen Jiang, Chao-Kai Wen, Shi Jin, and Jun Zhang
			\thanks{P. Jiang and S. Jin are with the National
				Mobile Communications Research Laboratory, Southeast University, Nanjing
				210096, China (e-mail: PeiwenJiang@seu.edu.cn; jinshi@seu.edu.cn).}
			\thanks{C.-K. Wen is with the Institute of Communications Engineering, National
				Sun Yat-sen University, Kaohsiung 80424, Taiwan (e-mail: chaokai.wen@mail.nsysu.edu.tw).}
			\thanks{J. Zhang is with the Department of Electronic and Computer Engineering,
				Hong Kong University of Science and Technology, Hong Kong (e-mail: eejzhang@ust.hk).}}
	
	\maketitle
	\pagestyle{empty}  
	\thispagestyle{empty} 

\begin{abstract}
Semantic communication, augmented by knowledge bases (KBs), offers substantial reductions in transmission overhead and resilience to errors. However, existing methods predominantly rely on end-to-end training to construct KBs, often failing to fully capitalize on the rich information available at communication devices. Motivated by the growing convergence of sensing and communication, we introduce a novel Position-Aided Semantic Communication (PASC) framework, which integrates localization into semantic transmission. This framework is particularly designed for position-based image communication, such as real-time uploading of outdoor camera-view images. By utilizing the position, the framework retrieves corresponding maps, and then an advanced foundation model (FM)-driven view generator is employed to synthesize images closely resembling the target images. The PASC framework further leverages the FM to fuse the synthesized image with deviations from the real one, enhancing semantic reconstruction. Notably, the framework is highly flexible, capable of adapting to dynamic content and fluctuating channel conditions through a novel FM-based parameter optimization strategy. Additionally, the challenges of real-time deployment are addressed, with the development of a hardware testbed to validate the framework. Simulations and real-world tests demonstrate that the proposed PASC approach not only significantly boosts transmission efficiency, but also remains robust in diverse and evolving transmission scenarios.

\end{abstract}

	\section{Introduction}

\IEEEPARstart{K}{nowledge} bases (KBs) are foundations of semantic communication systems, typically classified into local and shared KBs \cite{bao2011towards}. Different types of KBs are applied in various ways to optimize communication \cite{10554663}. For example,  KBs  derived from both tasks and channels were utilized in \cite{10628028} to manage semantic transmission. To address the challenge of limited explainability in current systems, a shared KB \cite{10318078} is used to integrate information and extract residual data. Similarly, in \cite{li2024end}, it was shown that shared KB-enabled generative semantic communication significantly reduces transmission loads by sending only relevant indices for image classification and generation tasks. Additionally, methods such as knowledge graphs and triples \cite{zhou2022cognitive} have proven effective in generating semantic symbols. By leveraging KBs to interpret transmission content, semantic communication can effectively save transmission bandwidth and improve transmission quality.

In contrast to conventional communication methods, the transmission modality in semantic communication plays a crucial role in its effectiveness. Visual modalities like images and videos, which convey richer information than text, have become a focal point in semantic communication research. For instance, a rate-adaptive coding technique was proposed in \cite{9953110} to enhance video transmission under limited bandwidth. A novel object-attribute-relation framework \cite{du2024object} enables efficient low bit-rate coding, improving video transmission performance. Semantic communication can also enhance visual quality in specific use cases. In video conferencing, as explored in \cite{9955991}, pre-shared static information allows real-time transmission of only essential keypoints. In another example, substation patrol inspections under poor channel conditions benefit from an emphasis on transmitting critical image details \cite{10329466}. Furthermore,   generative models are employed in \cite{lokumarambage2023wireless} to reconstruct images based on semantic features instead of raw pixels. Even multimodal content, as discussed in \cite{10622487}, can be efficiently transmitted using semantic techniques. These studies demonstrate the potential of semantic communication to transform the efficiency and quality of visual data transmission.

Meanwhile, the rise of foundation models (FMs), including large language models (LLMs) and diffusion models (DMs), has revolutionized the handling of different modalities, such as text and images. Integrating FMs into semantic communication systems can drastically improve performance and adaptability \cite{jiang2023large}. For instance, LLMs have been employed to coordinate multiple models within a communication system \cite{shen2023large}, and innovative semantic segmentation techniques have been developed using text-controlled content segmentation \cite{wang2023seggpt}. Diffusion models, on the other hand, can restore transmitted semantic features even with minimal data, such as object outlines \cite{grassucci2023generative}. LLMs are also being used to automate various communication modules like spectrum sensing and power allocation \cite{shao2024wirelessllm,xu2024semantic}, leveraging domain-specific knowledge and advanced knowledge alignment methods. In satellite communication, a novel FM-based framework \cite{jiang2024semantic} has been proposed to reduce bandwidth requirements and achieve precise feature recovery. These FM-based approaches provide robust solutions for comprehending complex transmission content and demonstrate remarkable adaptability in dynamic wireless communication environments.

Inspired by the advancements of integrated sensing and communications, particularly in position-aided communication techniques \cite{lohan2018benefits,li2017position}, we propose exploiting the position of transmitters to further enhance transmission efficiency by linking the position to transmission content. For example, in outdoor image transmission via camera sensors, position data can provide static information, such as buildings and streets, while fast-changing or dynamic objects like vehicles and signs become key elements in transmission. By incorporating FMs, we can effectively process semantic elements in three steps. First, the transmitted image is assessed for its position-based relevance. Then, a camera-view image synthesis is generated using the positional data. Finally, a robust image reconstruction model merges the positional and transmitted information to produce a high-quality final image. By incorporating these FM-based modules, we propose an adaptive transmission strategy that leverages an orthogonal frequency division multiplexing (OFDM) system to manage dynamic content and fluctuating environmental conditions, ensuring efficient and reliable communication.

Real-time implementation poses a significant challenge, particularly for FM-based semantic communication methods. While cutting-edge techniques \cite{zhu2023survey,huang2022context,song2023consistency} have reduced the complexity of FMs, making them implementable on mobile devices, lightweight models still require seconds to process. This delay may be acceptable in human-interaction scenarios, but it presents challenges for real-time physical layer modules in communication systems. In our study, we address this delay by proposing a robust and compact strategy to enhance transmission modules when KB updates are not available in real-time. More importantly, our method has been successfully implemented on a testbed \cite{ding2024adaptive}, comprising software-defined radio (SDR) and embedded signal processing modules that emulate mobile communication devices. Compared to existing testbeds for semantic communication \cite{yoo2023role}, our prototype is not only more portable but also emphasizes the lightweight and real-time capabilities of our approach. Additionally, we incorporate a high-performance server to provide edge computing resources, such as a base station (BS), enabling real-time FM integration. To the best of our knowledge, our testbed is one of the first to explore the integration of FMs in communication systems with server-side support.

The contributions of our study are as follows:
\begin{itemize}
\item \textbf{High performance:} By leveraging the position of camera sensors, transmission performance is significantly enhanced, especially with the use of FMs. Instead of transmitting all semantic elements, FMs, such as LLMs and DMs, use the position to retrieve relevant satellite maps and then synthesize camera-view images, reducing the transmission load to only the differences between transmitted and synthesized images. These elements are combined at the receiver, and a DM reconstructs a high-quality image.

\item \textbf{Adaptability:} In scenarios where transmitted images differ significantly from synthesized ones due to position-irrelevant content or outdated KBs, the required transmission bandwidth increases. Furthermore, under varying channel conditions, different bandwidth allocations are required to protect transmitted information. We propose an LLM-based strategy to adapt to these changes, ensuring robust performance across different environments.

\item \textbf{Hardware implementation:} FMs are typically unsuitable for mobile device implementation due to processing and transmission delays, hindering real-time performance. In our approach, compact modules are deployed locally to offload processing from the FM. The FM is only activated when bandwidth is insufficient, or performance degrades. The proposed method also tolerates KB mismatches caused by processing and transmission delays. Finally, the entire system is validated through over-the-air (OTA) testing.
\end{itemize}

The remainder of this paper is organized as follows. Section \uppercase\expandafter{\romannumeral2} describes the system model, including the semantic communication architecture and the impacts of introducing FMs. The proposed transmission framework is presented in Section \uppercase\expandafter{\romannumeral3}. Section \uppercase\expandafter{\romannumeral4} demonstrates the advantages of the proposed networks under various content and channel conditions. Section \uppercase\expandafter{\romannumeral5} details the implementation of the OTA testbed and discusses the results. Finally, Section VI concludes the paper.

\section{System Model}
\label{SystemModel}
In this section, we outline the semantic communication process that leverages KBs, focusing on both semantic coding and the OFDM transceiver. Additionally, we discuss the benefits and challenges associated with integrating FM-aided semantic communication.

%
%

\begin{figure*}
	\centering
	{\includegraphics[width=0.8\linewidth]{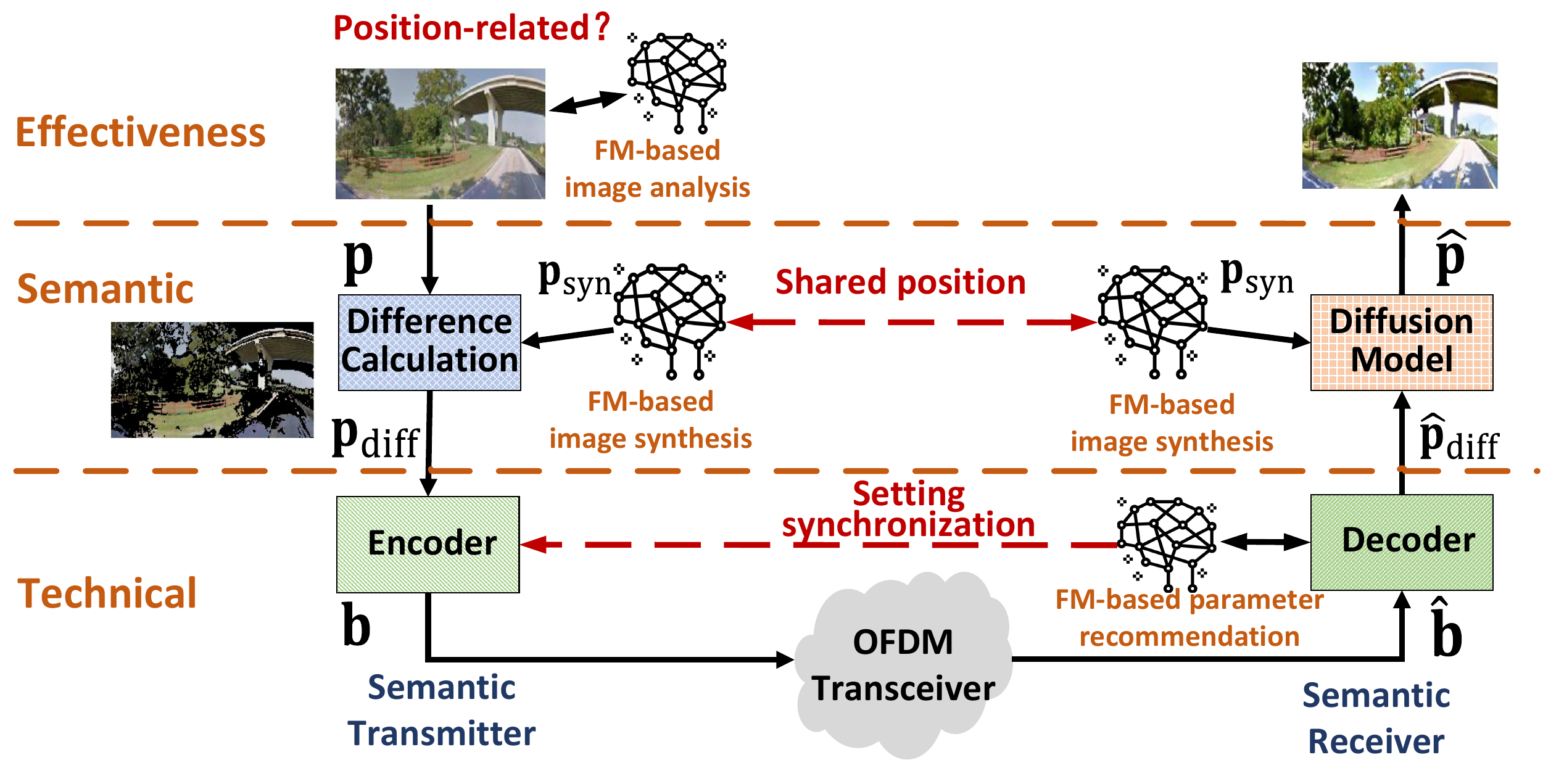}}
    \caption{Overview of the PASC framework.}
	\label{Overall}
\end{figure*}

\subsection{Semantic Communication in OFDM Systems} \label{SemanticCommunication&OFDM}
Recent advancements in deep learning, particularly with FMs, have enabled the efficient extraction and utilization of background knowledge in transmitted content. Initially, KBs were integrated implicitly through end-to-end training, where the knowledge was embedded within the trained model parameters. More recently, explicit KBs have been explored to enhance specific tasks. Various FMs, with their broad capabilities, are increasingly used to establish multi-level KBs \cite{10558822}.

Consider a semantic encoder, denoted as $\Theta_{\rm en}$, which utilizes KBs for transmitting an image  $\mathbf{p}$. The encoded codeword can be expressed as
\begin{equation}	
\mathbf{b}={\sf SC}_{\rm en}(\mathbf{p}; \Theta_{\rm en}),
\end{equation}
where ${\sf SC}_{{\rm en}}(\cdot)$ extracts the most relevant semantic features from the source content using the KB to form the codeword for transmission. Once the codeword $\mathbf{b}$ passes through a conventional transceiver, it is modulated into an OFDM symbol $\mathbf{x}$. The received signal in the frequency domain is then expressed as:
\begin{equation}	
	\mathbf{y}=\mathbf{h}\odot\mathbf{x}+\mathbf{z},	
\end{equation}
where $\odot$ represents element-wise multiplication, and $\mathbf{h}\in \mathbb{C}^{K \times 1}$ and $\mathbf{z}\in \mathbb{C}^{K \times 1}$ is the channel frequency response and the noise of the $K$ subcarriers. The received OFDM symbol can then be processed using the channel estimation result $\widehat{\mathbf{h}}$, yielding
\begin{equation}	
	\widehat{\mathbf{x}}=  \mathbf{y} \oslash \widehat{\mathbf{h}},
\end{equation}
where $\oslash$ denotes element-wise division. After demodulation, the estimated codeword $\widehat{\mathbf{b}}$ is obtained from $\widehat{\mathbf{x}}$. The image is then reconstructed by the semantic decoder ${\sf SC}_{\rm de}(\cdot)$ as
\begin{equation}	
	\widehat{\mathbf{p}}={\sf SC}_{\rm de}(\widehat{\mathbf{b}}; \Theta_{\rm de}),	
\end{equation}
where $\Theta_{\rm de}$ represents the KB at the receiver.

When designing the KBs for the semantic encoder-decoder, two common strategies are employed:
\begin{itemize}
\item \textbf{Implicit:} In this strategy, the knowledge is embedded in the trainable weights of the encoder and decoder, denoted as $\mathbf{W}_{\rm en}$ and $\mathbf{W}_{\rm de}$, respectively.  Thus, $(\Theta_{\rm en},\Theta_{\rm de})=(\mathbf{W}_{\rm en},\mathbf{W}_{\rm de})$.  The KBs at both the transmitter and receiver are shared through end-to-end training, with both sets of weights adjusted simultaneously. However, if the transmission scenario changes, the KB must be updated by fine-tuning or retraining the model.

\item \textbf{Explicit:} In certain scenarios, static elements within images (e.g., backgrounds or frequently appearing speakers in video conferencing) can be explicitly shared between the transmitter and receiver. For example, a shared image $\mathbf{p}_{\rm shared}$ can serve as part of the KB. In this case, the KB for the encoder is expressed as $\Theta_{\rm en}=(\mathbf{W}_{\rm en}, \mathbf{p}_{\rm shared})$, with a corresponding setup for the decoder. This approach provides a flexible and efficient method to adapt to new transmission scenarios by simply replacing the shared image $\mathbf{p}_{\rm shared}$.
\end{itemize}

\subsection{Benefits and Challenges of Integrating FMs}
FMs offer significant advantages for communication systems. Compared to traditional models, FMs demonstrate remarkable versatility in handling diverse content and dynamic transmission conditions, making them particularly well-suited for communication scenarios where both content and channel environments frequently change. Moreover, FMs contain extensive embedded knowledge, enabling a deeper understanding of the transmission content and facilitating more effective semantic feature extraction and reconstruction.

However, FMs are not inherently optimized for use in communication systems, where efficiency and speed are critical. A key limitation of FMs is their substantial processing time, which can span several seconds. Communication systems, on the other hand, require operations to be completed within milliseconds to avoid latency. While it is becoming increasingly feasible to run large FMs in the cloud, particularly for portable devices, this introduces additional delays due to the need for cloud interaction.

\section{Position-Aided Semantic Image Transmission }
\label{s3}
This section presents the framework of the proposed PASC system, which fully utilizes both transmitted image and position information. We describe the key components and adaptive methods, including detailed network architectures and the use of LLMs for parameter optimization. Additionally, we outline the challenges of implementing FM-based methods and offer recommendations for establishing a testbed for this approach.

\subsection{General Framework}
\label{s3a}

\begin{figure}
	\centering
	{\includegraphics[width=1.0\linewidth]{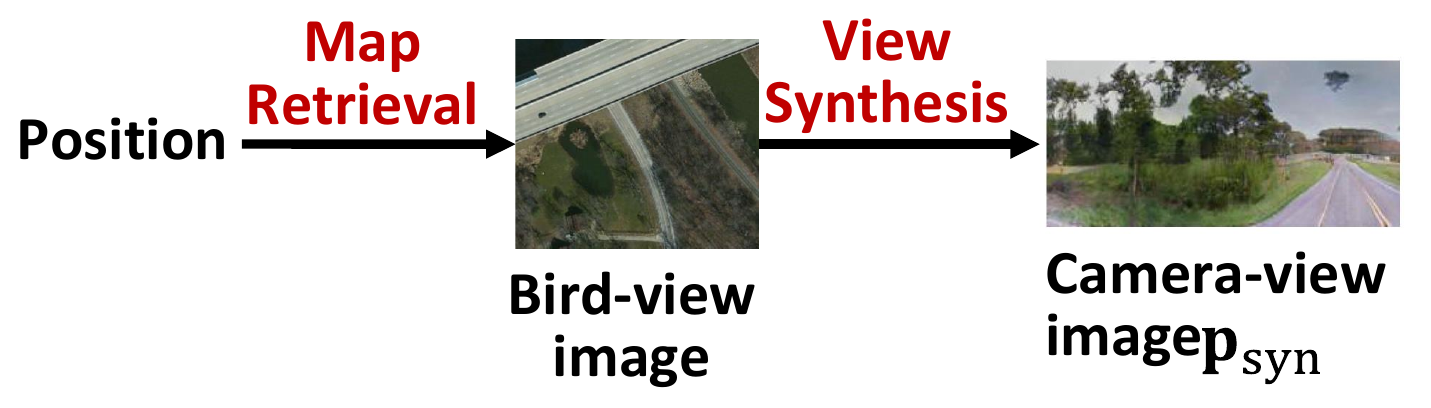}}
    \caption{Details of the FM-based image synthesis modules at the semantic level.}
	\label{FM-based_image_synthesis_modules}
\end{figure}

As depicted in Fig. \ref{Overall}, the transmission process in the PASC system is structured into three levels: effectiveness, semantic, and technical \cite{bao2011towards}, each corresponding to specific functions.
\begin{itemize}
\item The {\bf effectiveness level} focuses on analyzing the original and reconstructed images to ensure effective communication. Before using position information to enhance image transmission, it is important to confirm that the transmitted image is contextually relevant to the position. State-of-the-art LLM-based Visual Question Answering (VQA) applications are capable of understanding image content. For example, we can query, ``{\tt Is it an outdoor image?}'' If the position information is found to be useful, it must be made available to the receiver.

\item The {\bf semantic level} leverages FMs to utilize the transmitter's position as shared knowledge between the transmitter and receiver. FM-based image synthesis modules are deployed at both ends. As shown in Fig. \ref{FM-based_image_synthesis_modules}, the FM-based synthesis module accesses real-time local views from map services and retrieves a bird's-eye view image from an offline satellite map, based on the transmitter's known position. Using advanced view synthesis techniques \cite{toker2021coming,li2024sat2scene}, this bird's-eye view image is then converted into a camera-view image. In certain instances, map services may even provide direct ground-view images for specific locations.

\item The {\bf technical level} manages the physical transmission of data over wireless channels using semantic encoder-decoder modules and an OFDM transceiver, as described in Section \ref{SemanticCommunication&OFDM}. Traditional transmission parameters, such as the code rate, are adaptively adjusted by LLMs to optimize performance based on the content and transmission conditions.
\end{itemize}

In this framework, a DM-based generator is trained to convert the bird's-eye view image into a camera-view image. This synthesized image, referred to as $\mathbf{p}_{\rm syn}$, acts as a shared reference for both the transmitter and receiver, thus reducing the volume of data that needs to be transmitted. However, the synthetic image $\mathbf{p}_{\rm syn}$ will not be an exact replica of the real image captured by the transmitter. To address this, the transmitter computes the difference between the original and synthetic images, transmitting only the essential portions of the image, referred to as $\mathbf{p}_{\rm diff}$. The receiver then reconstructs the final image by combining the synthetic image with the received difference data.

The following subsections will provide detailed descriptions of the design and implementation of these modules.

\begin{figure*}
	\centering
	\subfigure[]{
		\label{En_De}
		{\includegraphics[width=0.95\linewidth]{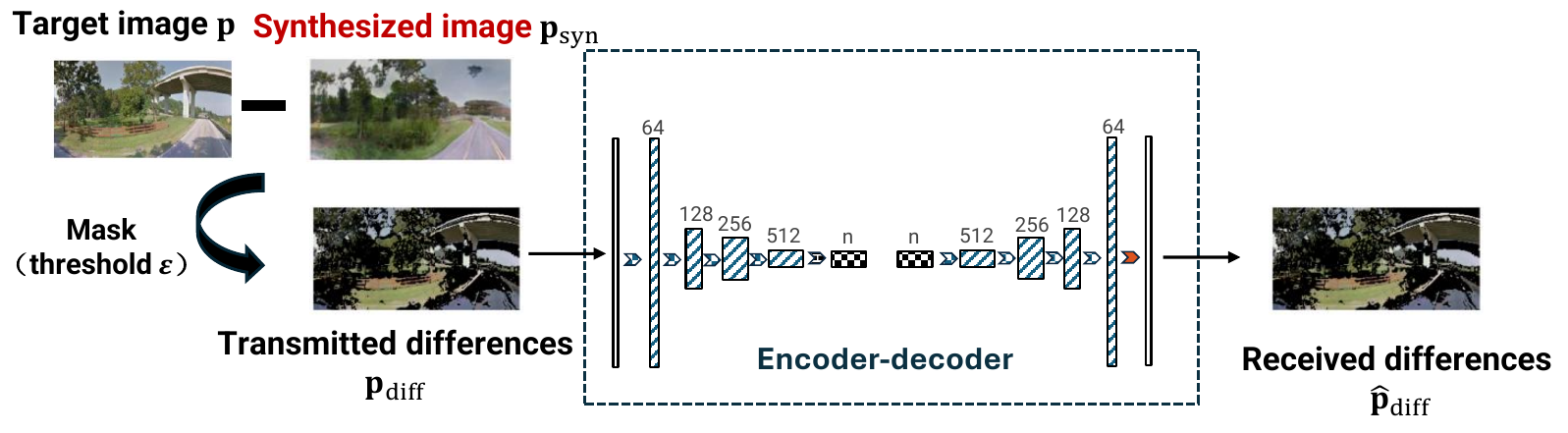}}}\\
	\subfigure[]{
		\label{DM}
		{\includegraphics[width=0.9\linewidth]{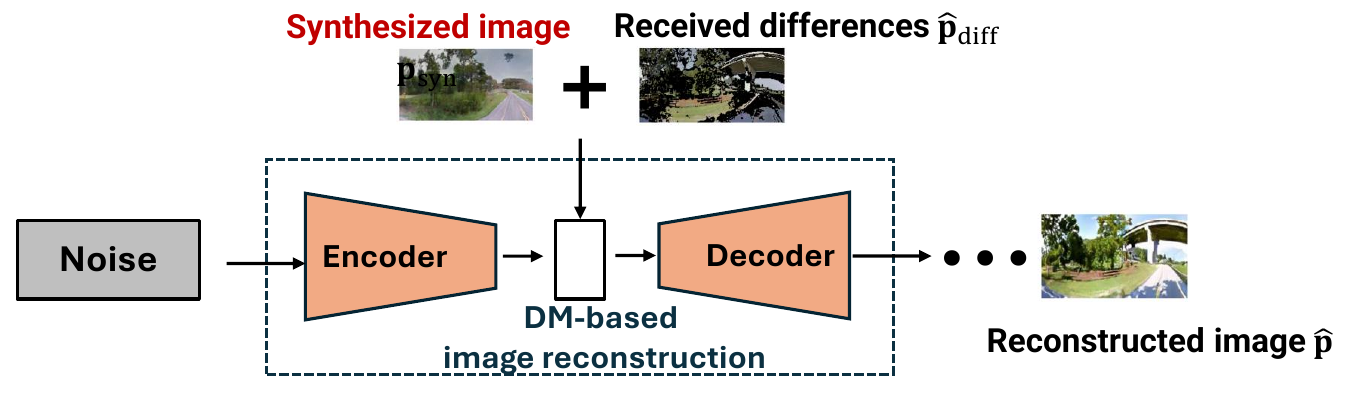}}}
	\caption{Detailed networks in the PASC framework. (a) Semantic encoder-decoder. (b) DM-based semantic reconstruction.}
	\label{Detailed network}
\end{figure*}

\subsection{Semantic Transceiver Design}
\label{s3b}

In the proposed system, a shared position is used to generate a synthetic camera-view image, $\mathbf{p}_{\rm syn}$, which typically captures most of the static objects and background in the scene. Therefore, only the differences between the original image and the synthetic image need to be transmitted. To further conserve bandwidth, minor differences that have little impact on the overall semantic meaning of the image are ignored, as they can be easily reconstructed through semantic reconstruction techniques. As illustrated in Fig. \ref{Detailed network}(a), the transmitted differences are defined as follows
\begin{equation}
		p_{{\rm diff},i,j}=\left\{
            \begin{array}{ll}
			p_{i,j}-p_{{\rm syn},i,j},&\sum_{k} |p_{i,j,k}-p_{{\rm syn},i,j,k}|> \varepsilon,\\
			0, & \text{otherwise}.
		\end{array}\right. \label{eq:def_varepsilon}
\end{equation}
In this equation, $p_{{\rm diff},i,j}$, $p_{i,j}$ and $p_{{\rm syn},i,j}$ represent the $(i,j)$-th pixels of the transmitted differences $\mathbf{p}_{\rm diff}$, the original image $\mathbf{p}$, and the generated synthetic image $\mathbf{p}_{\rm syn}$, respectively, with the last dimension $k$ corresponding to the RGB channels. For an image of size (256, 128, 3), the indices are $i \in \{ 1, \dots, 256\} $, $j \in \{ 1, \dots, 128\}$, and $k \in \{ 1, 2, 3\}$. Note that the pixel values are normalized to the range $[-1, 1]$. The threshold $\varepsilon$ controls how much of the information is transmitted, where all differences are transmitted when $\varepsilon=0$.

The next step involves encoding the transmitted difference data $\mathbf{p}_{\rm diff}$ into a codeword $\mathbf{b}$ using an encoder. The encoder consists of four convolutional blocks and a quantization layer, as described by
\begin{equation}
	\mathbf{b}=f_{\rm en}(\mathbf{p}_{\rm diff};\, \mathbf{W}_{\rm en}),
\end{equation}
where $ f_{\rm en}(\cdot) $ represents the encoding process. The encoder's architecture includes a $ 13 \times 13 $ convolution operation with 64 channels, followed by a $ 4 \times $ downsampling layer. The subsequent blocks use $7 \times 7$, $5 \times 5$, and $3 \times 3$ convolution operations with 128, 256, and 512 channels, respectively, each with a $2 \times$ downsampling layer. ReLU is used as the activation function in these blocks. The quantization layer flattens the input values, followed by a dense layer with a Tanh activation function, which outputs a number of values equal to the transmitted bits. A hard decision criterion is then applied to convert these values into the binary sequence $\mathbf{b}$, with gradients rewritten for end-to-end training.

During the OFDM transmission, the received binary sequence $\widehat{\mathbf{b}}$ is processed through a dense layer with a ReLU activation function, which restores it to the same dimensions as the convolutional blocks in the encoder. The decoder then mirrors the encoder's architecture, using four convolutional blocks with upsampling layers. Finally, a $3 \times 3$ convolutional layer with a Tanh activation function outputs the reconstructed difference image $\widehat{\mathbf{p}}_{\rm diff}$. This process is described as
\begin{equation}
	\widehat{\mathbf{p}}_{\rm diff}=f_{\rm de}(\widehat{\mathbf{b}};\, \mathbf{W}_{\rm de}),
\end{equation}
where $f_{\rm de}(\cdot)$ represents the decoding process.

This semantic transceiver is designed as a joint source-channel coding system, which is trained in an end-to-end manner. Due to the non-differentiable nature of the traditional OFDM transmission process under multiplicative wireless channels, the transmitted sequence $\mathbf{b}$ is trained with a set bit error rate (BER) of 0.01. This bit error rate is modeled as a binary symmetric channel, denoted as  $h_{\rm B}(\cdot)$. The training process is formulated as
\begin{equation}
	\left(\widehat{\mathbf{W}}_{{\rm en}},\widehat{\mathbf{W}}_{{\rm de}}\right) =\mathop{\arg\min}\limits_{\mathbf{W}_{{\rm en}},\mathbf{W}_{{\rm de}}} \left\|\mathbf{p}_{\rm diff}-f_{\rm de}(h_{\rm B}(f_{\rm en}(\mathbf{p}_{\rm diff}))) \right\|^2.
\end{equation}
Once trained, the semantic encoder-decoder system is integrated with the OFDM transceiver and is ready for the efficient transmission of the difference image $\mathbf{p}_{\rm diff}$.

\subsection{DM-based View Synthesis and Semantic Reconstruction}
DMs, as a prominent type of FM, are widely used in tasks such as image generation and inpainting. Conditional DMs, in particular, have demonstrated their ability to dynamically incorporate available conditions to reconstruct high-quality images, either from noisy input images or key semantic elements.


As illustrated in Fig. \ref{Detailed network}(b),  this study employs a standard conditional DM \cite{dhariwal2021diffusion}, where encoded conditions are integrated into the diffusion process. Each step in the diffusion process uses a UNet to progressively denoise the input image, guided by the encoded conditions $\mathbf{C}$. This can be described by the following expression
\begin{equation}
	\mathbf{p}^{(t)}=f_{\rm UNet}(\mathbf{p}^{(t-1)},\mathbf{C}),
\end{equation}
where $\mathbf{p}^{(t)}$ is the output image at the $t$-th step and $f_{\rm UNet} $ represents the denoising process. The initial input at the first step is a Gaussian noise sample, denoted as $\mathbf{p}^{(0)}=\mathbf{z}$, which has the same dimensions as the target output image. After completing $T$ steps of diffusion, the final generated image is obtained, and the entire DM process can be expressed as
 \begin{equation}
 	\mathbf{p}^{(T)}={\sf DM}(\mathbf{z},\mathbf{C}).
 \end{equation}
For simplicity, when referring to the final output image, we omit the superscript $^{(T)}$ from $\mathbf{p}$ in subsequent discussions.

Within this conditional DM framework, we design two critical modules:
\begin{enumerate}
  \item \textbf{View Synthesis:} Given that the camera's position and viewing direction, denoted as $\Theta_{\rm pos}$, are known through sensing and shared from the transmitter to the receiver, a corresponding bird's-eye view image can be retrieved and processed using an offline mapping service, represented by ${\sf MAP}(\cdot)$. For simplicity, we assume the transmitter is positioned at the center of the image and faces upward. The processed map, which incorporates both positional and directional information, serves as the input condition for the DM-based synthesis module. The synthetic image is generated as follows
  \begin{equation}
 	\mathbf{p}_{\rm syn}={\sf DM}_{\rm syn}(\mathbf{z},{\sf MAP}(\Theta_{\rm pos})).
 \end{equation}

 \item \textbf{Semantic Reconstruction:}
 For image reconstruction, the received difference image $\widehat{\mathbf{p}}_{\rm diff}$ and the synthetic image $\mathbf{p}_{\rm syn}$ are combined to form the condition $\mathbf{C}=[\widehat{\mathbf{p}}_{\rm diff},\mathbf{p}_{\rm syn}]$. The reconstructed image is generated by the DM-based reconstruction module, expressed as
 \begin{equation}
 	 \widehat{\mathbf{p}}={\sf DM}_{\rm rec}(\mathbf{z},[\widehat{\mathbf{p}}_{\rm diff},\mathbf{p}_{\rm syn}]).
 \end{equation}
 During training, the SNR is varied from 0 to 10 dB when generating $\widehat{\mathbf{p}}_{\rm diff}$. This ensures that the trained DM can handle dynamic transmission scenarios, making the reconstruction process robust to different channel conditions.
\end{enumerate}

\subsection{Adaptation to Content and Environment}

\begin{figure*} 
	\centering
    \includegraphics[width=0.8\linewidth]{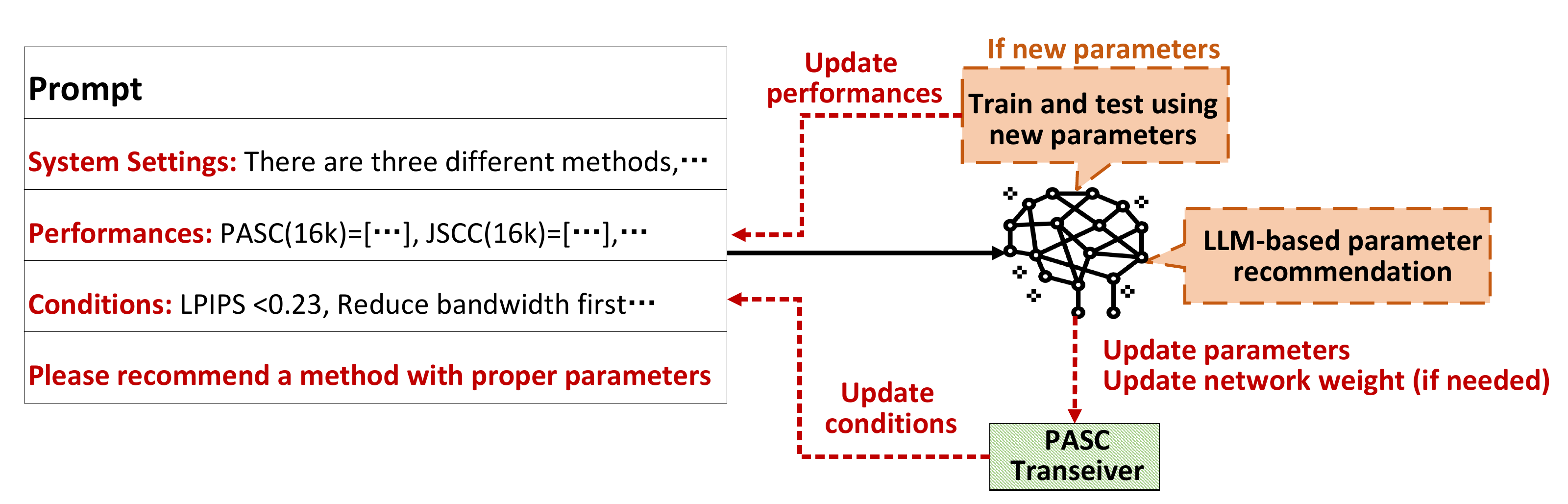}
    \caption{Process of the LLM-based adaptive method. The LLM recommends system parameters based on the input information, and these recommended parameters are used to further update the reference information.}
	\label{LLM}
\end{figure*}

Pretrained networks, as described earlier, often struggle to maintain optimal performance when faced with varying transmission content and changing channel conditions. Moreover, fixed-length transmission codewords are not always efficient, as they fail to balance bandwidth consumption with image quality. To address these challenges, we propose methods that adapt to both content and environmental variations by leveraging FMs and compact models.

\textbf{JSCC Method for Position-Independent Content:}
When the transmission content is independent of the transmitter's position, a robust transmission method is required. In such cases, encoder-decoder networks can be trained to transmit the entire image,  $\mathbf{p}$. The received image, denoted as $\widehat{\mathbf{p}}_{\rm JSCC}$, is reconstructed as follows
\begin{equation}
\widehat{\mathbf{p}}_{\rm JSCC}=f_{\rm de}(h_{\rm B}(f_{\rm en}(\mathbf{p}))),
\end{equation}
where the network weights are retrained by minimizing the following loss
\begin{equation}
	\left(\widehat{\mathbf{W}}_{{\rm en}},\widehat{\mathbf{W}}_{{\rm de}}\right)
    =\mathop{\arg\min}\limits_{\mathbf{W}_{{\rm en}},\mathbf{W}_{{\rm de}}} \left\|\widehat{\mathbf{p}}_{\rm JSCC}-f_{\rm de}(h_{\rm B}(f_{\rm en}(\mathbf{p}))) \right\|^2.
\end{equation}
To further demonstrate the effectiveness of this method, the entire image can also be transmitted and reconstructed using a DM. The received image $\widehat{\mathbf{p}}_{\rm JSCC}$ is used as a condition to generate the reconstructed image, referred to as $\widehat{\mathbf{p}}_{\rm JSCC+DM}$.

\textbf{Adjusting Threshold for Varying  Content:}
The threshold $\epsilon$ used in (\ref{eq:def_varepsilon}) plays a key role in controlling the transmitted differences. When bandwidth is limited, a higher value of $\epsilon$ transmits only the most significant differences. For example, when $\epsilon = 0.4$, approximately 60\% of the pixels in $\mathbf{p}_{\rm diff}$ are zeros, and this proportion increases to 90\% when $\epsilon = 1$. In this study, the default value of $\epsilon = 0.4$ is used for the PASC methods.

\textbf{Multiple Code Rates for Varying Bandwidth:}
To accommodate varying bandwidth requirements, multiple encoder-decoder pairs are trained with different configurations. Specifically, the quantization layer in the encoder and the entire decoder are retrained under various settings, with adjustments to both the content threshold $\epsilon$ and the encoder's output dimension. These retrained parameter sets are saved and labeled accordingly, such as PASC(1k, $\epsilon=0.4$), where ``1k'' refers to one thousand transmitted bits, and $\epsilon = 0.4$ is the threshold. Similarly, encoder-decoders trained for transmitting the entire image under different bandwidth conditions are labeled as JSCC(1k) or JSCC+DM(1k), depending on whether DM is used.

Once these parameter sets are saved, it is essential to develop an adaptive strategy to dynamically adjust to different content and environmental conditions. As discussed in Section \ref{s3a}, the LLM's VQA capability can be used to differentiate content. Additionally, the LLM can assist in selecting the most appropriate parameter set from the available options based on the transmission requirements.
The strategy's prompt consists of three key components: system settings, performance data, and conditions, as illustrated in Fig. \ref{LLM}.
{\bf System settings} provide basic task information, such as: ``{\tt There are three different methods: JSCC, JSCC+DM, and PASC ...}''.
{\bf Performance data} lists simulation results of each method's effectiveness, along with their computational requirements under different conditions.
{\bf Conditions} prompt the LLM to select the best method and corresponding settings for a given scenario. If existing settings are insufficient, the LLM can recommend new configurations, which are then trained and tested. Once validated, these new settings are added as available options.

By utilizing LLM-based recommendations and adaptive strategies, the proposed transmission framework can respond to varying requirements and even handle new ones through automatic retraining. However, the complexity of FMs can occasionally introduce errors, making robustness and real-time implementation important topics for further investigation.

\subsection{Robustness and Real-time Solutions for Hardware Implementation}

High-performance servers typically refer to BSs and edge computing nodes in communication systems, which provide substantial computational resources. In the uplink transmission scenario of this study, the receiver (BS) has sufficient computational capacity locally, and thus the receiver can directly use FMs.  However, the transmitter (camera sensor) relies on the edge server and thus cannot use FMs often considering the interaction and process delays. This entire transmission framework involves multiple devices, and maintaining real-time interaction between these devices is challenging. The runtime of complex FMs is typically measured in seconds, while the duration of a standard OFDM frame is in milliseconds (ms). Consequently, FMs may not be able to make decisions in time, resulting in outdated information. Furthermore, FMs may also make mistakes, and the combination of outdated information and errors can lead to \emph{mismatched} KBs provided by FMs. For instance, the synthetic image may have a different background from the transmitted image due to processing delays or erroneous judgments.

To address the mismatch caused by FMs, ensuring the robustness of the proposed methods is critical for maintaining transmission performance. The following discussion focuses on issues related to robustness and real-time operation.

\textbf{Transmitter:} The transmitter cannot frequently communicate with the server due to the limited communication and computational resources available at edge computing nodes, especially when a single edge computing node serves numerous sensor terminals. Fortunately, the scenario for the camera sensor does not change often, as most camera sensors are either static or move slowly. When the transmitted image is position-related, the synthetic camera-view image can be generated and remain effective for a certain period. For example, when the camera moves slightly within the same bird's-eye view image, the bird's-eye view can be recropped to center the camera for view synthesis. In this case, without server assistance, local view synthesis can be performed using lightweight UNets trained for specific ranges of locations, providing performance close to that of DM within those ranges.

\textbf{Receiver:} The receiver at the BS, equipped with ample computational resources, performs well once the transmitter's current position is known. However, it is crucial to promptly update information such as whether the transmitted image is position-related, the current position, and the satellite map to maintain the consistency of the shared KB with the transmitter. With the advancement of state-of-the-art acceleration techniques \cite{ma2024deepcache} for FMs, processing signals using FMs at BSs will become increasingly feasible.

From the above analysis, performance loss usually stems from untimely updates. For example, if the satellite map is mismatched with the current transmitted image or if the camera-view synthesis is inaccurate due to an outdated local UNet, transmission performance suffers. In cases where the camera sensor moves indoors but there is a delay before the devices become aware of the change through LLM judgment, a mismatch arises between $\mathbf{p}_{\rm syn}$ and $\mathbf{p}$, altering the distribution of the difference $\mathbf{p}_{\rm diff}$ and degrading encoder-decoder performance. Furthermore, DM-based reconstruction may be misled by mismatched $\mathbf{p}_{\rm syn}$. To address these issues, the proposed methods should be robust to mismatched conditions. JSCC methods can be employed as a fallback option that does not rely on mismatched KBs. A rapid assessment using an encoder and several dense layers is proposed to distinguish $\mathbf{p}_{\rm diff}$ and determine whether $\mathbf{p}$
corresponds to $\mathbf{p}_{\rm syn}$.

The proposed methods leverage the strengths of FMs in tasks such as scenario distinction, KB establishment, and image reconstruction. At the same time, innovative strategies are designed to mitigate the drawbacks caused by FM errors and high complexity. In summary, this flexible, position-aided semantic communication framework is highly competitive, and its performance under various conditions is demonstrated in the following sections.

\begin{figure*} 
	\centering
	\subfigure[]{
		\label{Diff_RIS_ACC}
		{\includegraphics[width=0.48\linewidth]{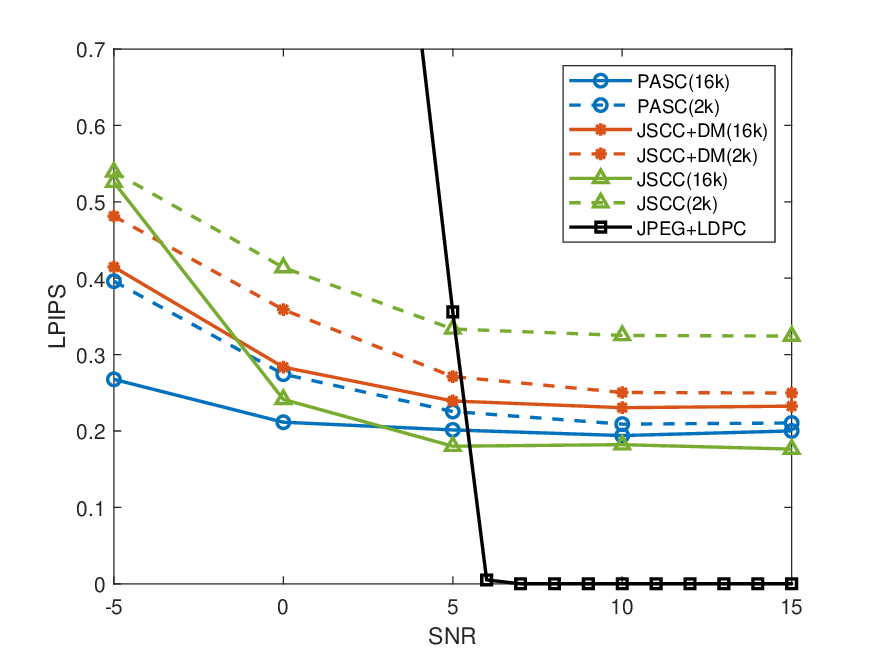}}}
	\subfigure[]{
		\label{Diff_RIS_MSE}
		{\includegraphics[width=0.48\linewidth]{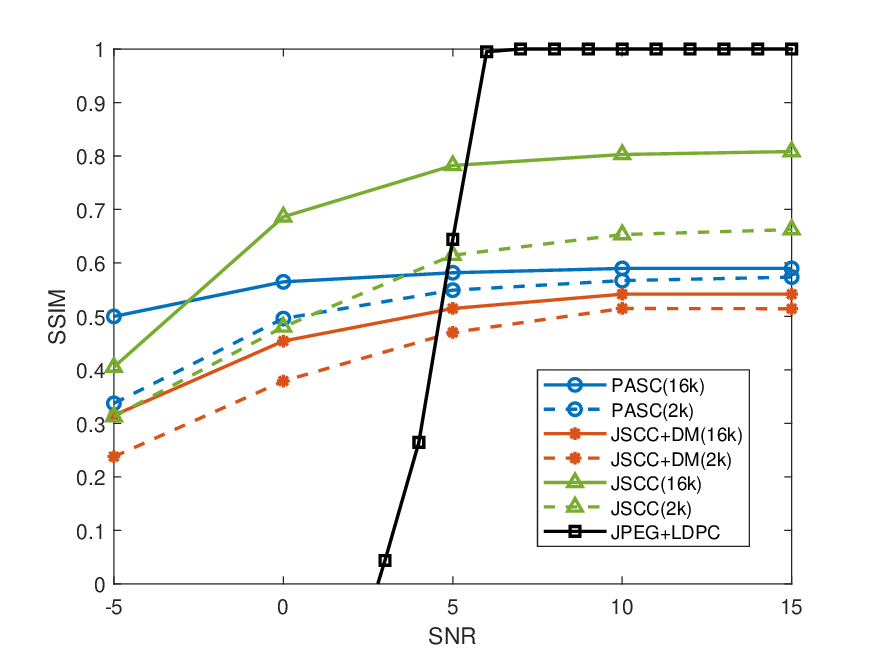}}}
    \caption{Performance comparison of the competing methods across different bandwidths. (a) LPIPS metric. (b) SSIM metric.}
	\label{SecA_per}
\end{figure*}

\begin{figure*} 
	\centering
	
	\subfigure[Target image]{
		\includegraphics[width=0.23\linewidth]{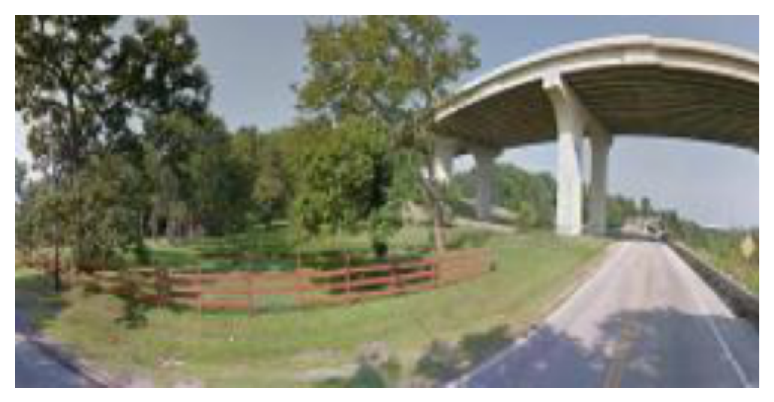}}
	\subfigure[JSCC(2k), SNR=$5$ dB]{
		\includegraphics[width=0.23\linewidth]{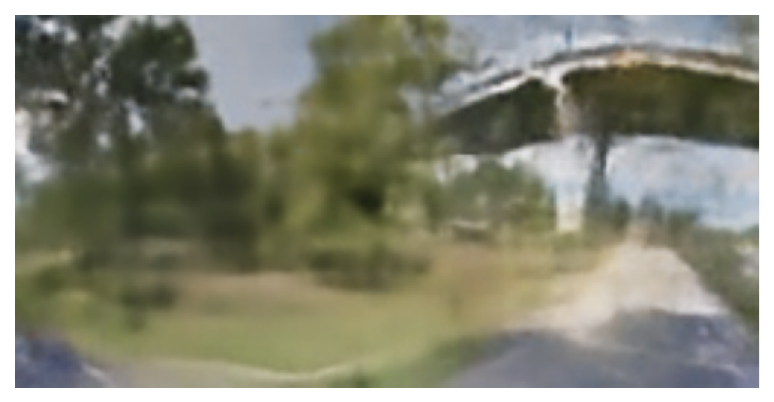}}
	\subfigure[JSCC+DM(2k), SNR=$5$ dB]{
		\includegraphics[width=0.23\linewidth]{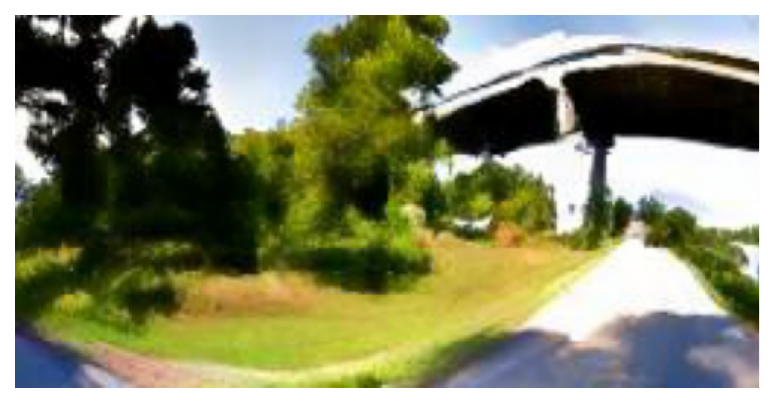}}
	\subfigure[PASC(2k), SNR=$5$ dB]{
		\includegraphics[width=0.23\linewidth]{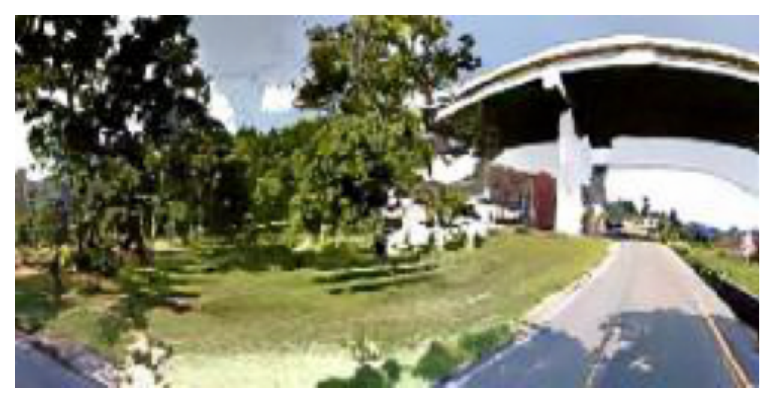}}\\
	
	\subfigure[Synthesized image]{
		\includegraphics[width=0.23\linewidth]{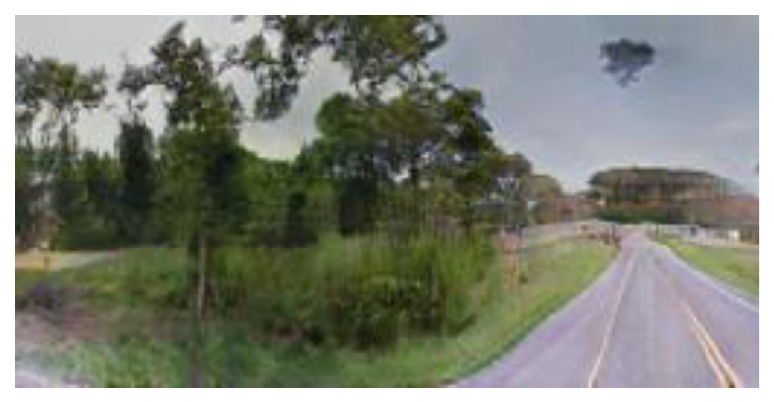}}
	\subfigure[JSCC(2k), SNR=$-5$ dB]{
		\includegraphics[width=0.23\linewidth]{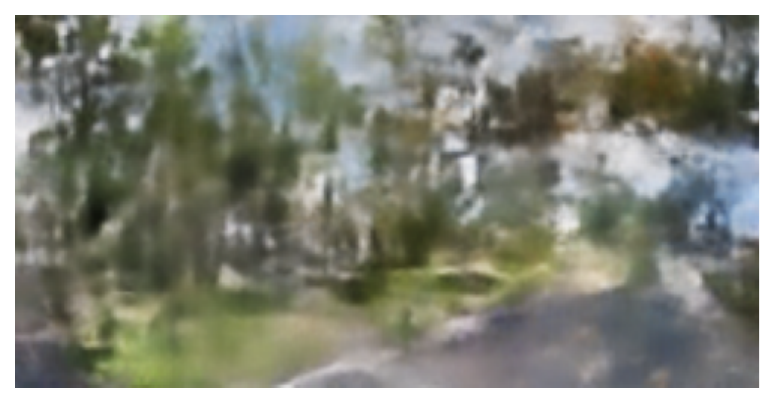}}
	\subfigure[JSCC+DM(2k), SNR=$-5$ dB]{
		\includegraphics[width=0.23\linewidth]{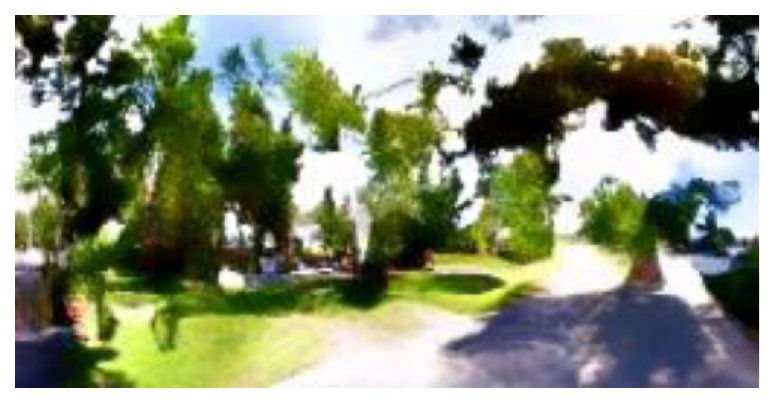}}
	\subfigure[PASC(2k), SNR=$-5$ dB]{
		\includegraphics[width=0.23\linewidth]{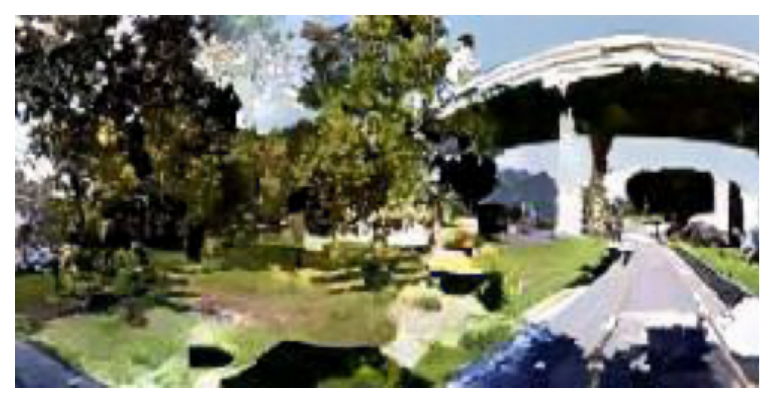}}
	\caption{Examples of the competing methods. (a) and (e) represent the target and synthesized images. (b) and (f), (c) and (g), and (d) and (h) correspond to the JSCC, JSCC+DM, and PASC methods with SNR=5 dB and SNR=-5 dB, respectively, each using two thousand transmission bits.}
	\label{EX_A}
\end{figure*}

\section{Simulation Results}
\label{s5}
After introducing the simulation settings, we compare the proposed PASC method with competing methods under various channel conditions. We also analyze the impact of different settings and demonstrate the effectiveness of the proposed adaptive strategy. Finally, we discuss the robustness of the different methods when mismatches occur.

\subsection{Simulation Settings}

The SUI-5 channel model is used, with three multipaths and a delay spread of 10. A conventional OFDM transceiver is simulated. Each OFDM symbol utilizes 256 subcarriers, with 25 allocated for pilot signals, 125 for data transmission, and the remaining subcarriers reserved for guard bands. The OFDM frame length is 13 ms, and each frame contains 40 OFDM symbols. The cyclic prefix (CP) length is 64, and QPSK modulation is used.

The CVUSA \cite{workman2015wide} and KITTI \cite{geiger2012we} datasets are used, containing more than ten thousand images from different areas, along with their positions and satellite maps. The training and testing set ratio is 10:1. All camera-view images are cropped to $256\times128$, and their corresponding satellite maps are processed to the same size, placing the camera at the center. The DM-based view synthesis is trained using data from all areas, while UNet-based view synthesis is trained separately for different areas.

Among the competing methods, the conventional approach consists of JPEG source coding and LDPC channel coding. The average bandwidth consumption of this JPEG+LDPC method is 35k bits, with an LDPC code rate of 1/2. The bandwidth settings for the PASC and JSCC methods are lower than those for the conventional method. JSCC has the lowest complexity among neural network-based methods since it does not use FMs. PASC and JSCC+DM have similar transmission complexity, but JSCC+DM does not require FMs for sharing KBs.

Three main metrics are used to explain the quality of the restored images.
\begin{itemize}
\item \textbf{SSIM:} SSIM evaluates the structural similarity between image patches \cite{hore2010image}. It is more robust than PSNR and is widely used as an image quality metric.

\item \textbf{LPIPS:} LPIPS is commonly used as a regularization method in computer vision network training. It calculates the sum of MSEs between the estimated and true images across different layers of a pretrained network, such as VGG, representing feature similarity. The LPIPS metric from \cite{zhang2018unreasonable} is used here.

\item \textbf{CLIP-I:} CLIP-I \cite{ruiz2023dreambooth} measures the average pairwise cosine similarity between the CLIP embeddings of generated and real images.

\end{itemize}

\subsection{Performance Under Different Channel Conditions}

In this subsection, we describe the impact of varying SNRs and bandwidth on the competing methods. Additionally, visual examples are provided to demonstrate the superiority of the proposed methods, particularly under challenging channel conditions.

As shown in Fig. \ref{SecA_per}(a), the LPIPS performance of the conventional JPEG+LDPC method is optimal when the SNR is high. However, as SNR decreases, the performance of this method significantly deteriorates once transmission errors exceed its correction capacity. In contrast, semantic methods consistently perform well under low SNR conditions. Among these methods, JSCC(16k) achieves the best performance when SNR $\geq$ 5 dB, as the bandwidth is sufficient for this relatively simple network. Generative FMs in JSCC+DM(16k) and PASC(16k) do not offer further enhancements under these conditions. However, when SNR $<$ 0 dB, JSCC(16k) performance declines rapidly, whereas JSCC+DM(16k) and PASC(16k) continue to effectively correct more semantic errors. When bandwidth is limited to 2k bits, PASC(2k) demonstrates clear advantages due to its minimal bandwidth requirement for transmitting differences.

Fig. \ref{SecA_per}(b) shows the comparative results using the SSIM metric. Notably, the conventional JPEG+LDPC method experiences a sharp decline as SNR decreases. JSCC methods generally outperform other semantic approaches, as generative models can sometimes alter the structural information of the image. The PASC method outperforms JSCC+DM because the synthetic image at the receiver preserves more structural information than direct generation from the received image. When SNR is below 0 dB, PASC(2k) outperforms JSCC(2k) because JSCC's entire transmitted image becomes corrupted under extremely poor conditions.

Several example transmissions using different methods are compared in Fig. \ref{EX_A}. Some objects, such as the bridge in Fig. \ref{EX_A}(a) and (e), are difficult to generate from the satellite map. This indicates that PASC must compensate for the differences between (a) and (e) through transmission. At $5$ dB SNR, JSCC(2k) can restore a noisy image, and the received image of JSCC+DM(2k) is clearer, enhanced by DM over JSCC's noisy result. PASC(2k) provides more details than the other methods at $5$ dB SNR. At $-5$ dB SNR, JSCC(2k) produces a broken image, and JSCC+DM(2k) can only infer the contents. PASC(2k), however, still performs relatively well under $-5$ dB SNR, although some noise-induced artifacts appear in the image.

In conclusion, the proposed PASC method effectively reduces bandwidth requirements and performs well under low SNR conditions. This improvement is achieved through position knowledge and generative FMs. Moreover, the advantages of different methods can be adaptively exploited, as discussed in the following section.

\begin{figure} 
	\centering
	\subfigure[]{
		
		{\includegraphics[width=0.98\linewidth]{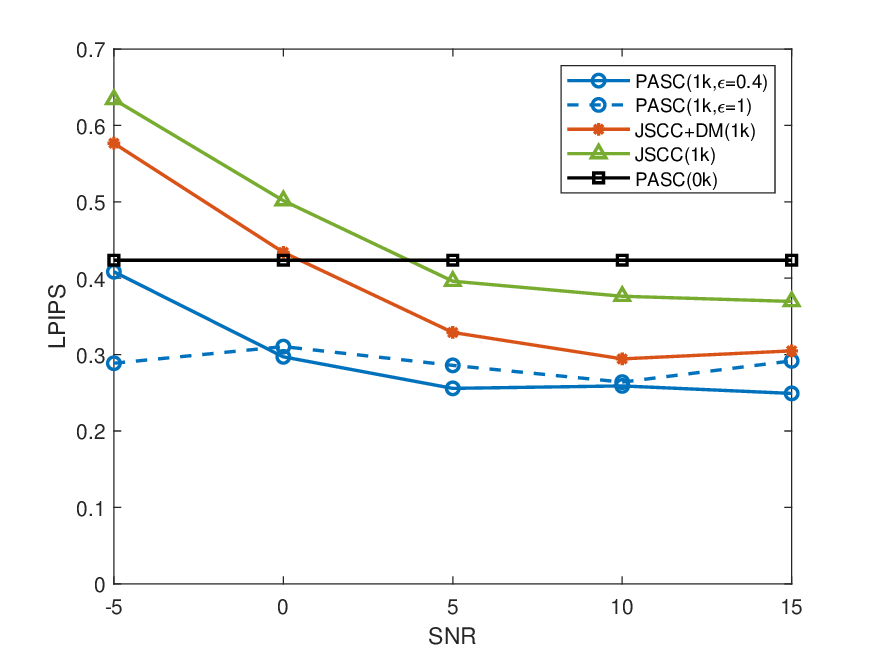}}}\\
	\subfigure[]{
		{\includegraphics[width=0.98\linewidth]{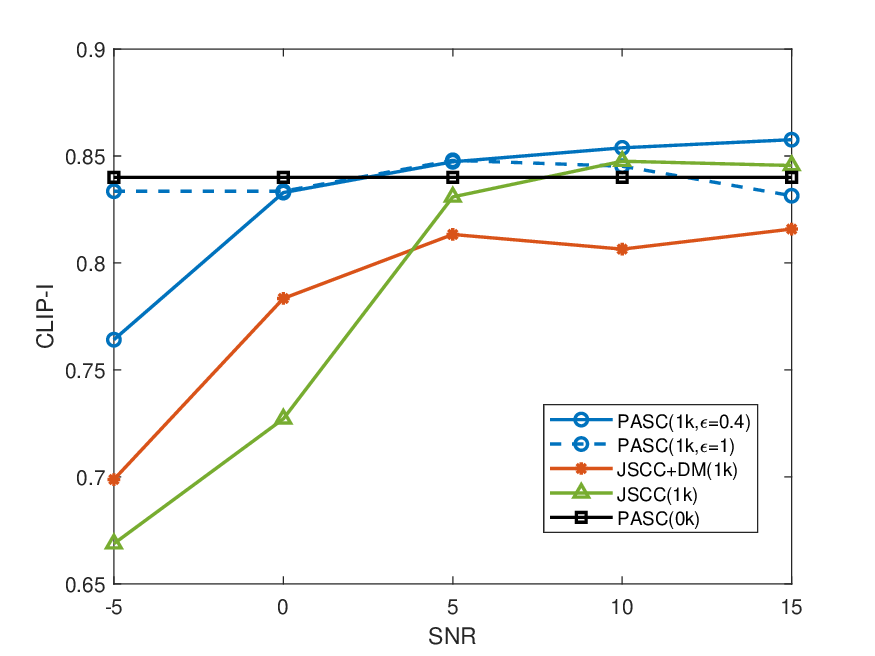}}}
    \caption{Performance of the competing methods under extremely low bandwidth conditions with varying thresholds. (a) LPIPS metric. (b) CLIP-I metric.}
	\label{SecB_per}
\end{figure}

\subsection{Impact of Different Settings and Adaptive Strategy}
In this subsection, we first compare the proposed PASC methods under different settings. We then discuss the benefits of using an LLM to recommend the choice of settings.

Fig. \ref{SecB_per}(a) shows the LPIPS performance of the competing methods using smaller bandwidth compared to Fig. \ref{SecA_per}(a). As a result, the performance gaps between PASC(1k, $\epsilon=0.4$), JSCC(1k), and JSCC+DM(1k) become more pronounced. Notably, PASC(0k) represents an extreme case where no information is transmitted, resulting in the received differences being purely noise. In this case, the DM cannot extract useful information from the received image and only generates an image similar to the synthetic image $\mathbf{p}_{\rm syn}$. When SNR $<$ 0, both JSCC(1k) and JSCC+DM(1k) perform worse than PASC(0k), indicating that these JSCC methods convey less information than the position-based $\mathbf{p}_{\rm syn}$. The performance of PASC(1k, $\epsilon=0.4$) is similar to PASC(0k) at $-5$ dB SNR, as the transmitted information cannot be effectively restored under such poor conditions. PASC(1k, $\epsilon=1$) outperforms PASC(1k, $\epsilon=0.4$) by transmitting fewer but more significant differences, thus preserving more bandwidth to mitigate transmission noise. However, this comes at the cost of sacrificing detailed differences, resulting in poorer performance compared to PASC(1k, $\epsilon=0.4$) when SNR $>$ 0 dB.

Fig. \ref{SecB_per}(b) evaluates the generated images using the CLIP-I metric, which is based on the features extracted by the CLIP network that connects image features with textual descriptions. Therefore, this metric prioritizes the similarity of text descriptions over detailed differences between images. The simulation results indicate that PASC methods consistently perform well by sharing synthesized images based on position information. Specifically, PASC(0k) shows high similarity between the generated and transmitted images using only position information and FMs. When SNR $>$ 0 dB, PASC(1k, $\epsilon=0.4$) and PASC(1k, $\epsilon=1$) perform slightly better than PASC(0k) due to the successful reception of additional details. However, for SNR $\leq$ 0 dB, PASC(1k, $\epsilon=0.4$) performs significantly worse than PASC(0k) as transmission errors introduce artifacts that negatively affect the text descriptions. PASC(1k, $\epsilon=1$) performs slightly worse than PASC(0k) due to fewer differences being transmitted, which better preserves the overall image structure.

\begin{table} 
	\centering	
	\scriptsize
    \caption{LPIPS performance of different settings recommended by LLM. The default threshold $\epsilon$ is 0.4.}
	
	\begin{tabular}{clllll}
		\toprule
		& SNR&$-5$ dB&0 dB &$5$ dB&$10$ dB  \\\midrule
	\multirow{2}{*}{\makecell{Example 1\\ (Complexity first)} }&Choose&\makecell{PASC\\ 16k} &\makecell{PASC\\ 16k}& \makecell{JSCC\\ 16k}&\makecell{JSCC\\ 16k}\\ \cmidrule{2-6}
	 &Performance&0.2676& 0.2115&0.1801&0.1821\\ \midrule
\multirow{2}{*}{\makecell{Example 2 \\(Bandwidth first)} }&Choose	&\makecell{PASC\\ 16k}& \makecell{PASC\\ 16k} &\makecell{JSCC\\ 16k}&\makecell{PASC\\ 2k}\\ \cmidrule{2-6}
	&	Performance &0.2676& 0.2115&0.2090&0.2105\\ \midrule
	\multirow{2}{*}{\makecell{Example 3\\ (Bandwidth first)} }&Recommend&\makecell{PASC\\ $\epsilon$=1, 8k}& \makecell{PASC\\ 16k} &\makecell{PASC\\ 8k}&\makecell{PASC\\ 2k}\\ \cmidrule{2-6}
	&Performance&0.2430&  0.2115&0.2259&0.2105 \\
		
		\bottomrule
	\end{tabular}
	\label{Metric1}
\end{table}

\begin{figure*} 
	\centering
	\subfigure[]{
		
		{\includegraphics[width=0.42\linewidth]{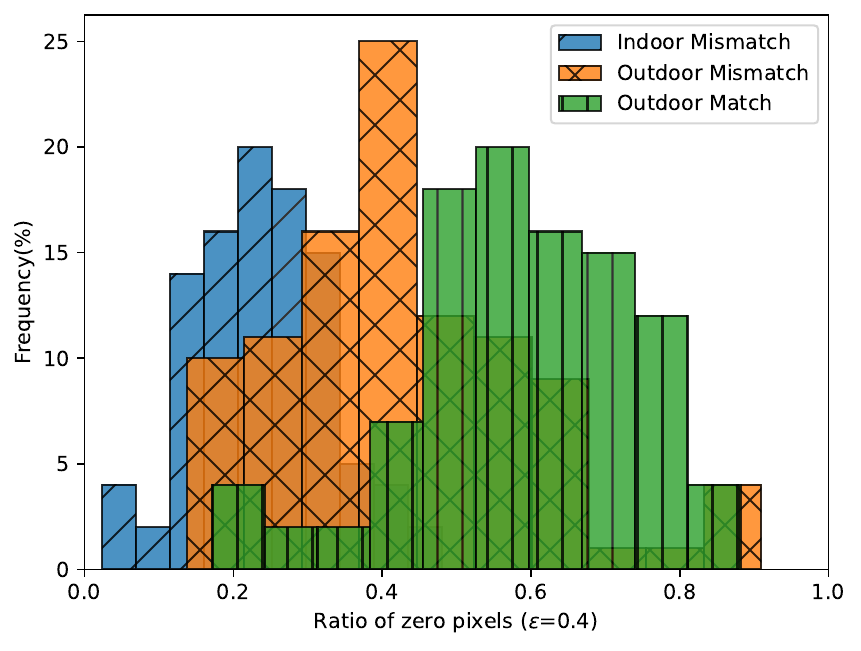}}}
	\subfigure[]{
		{\includegraphics[width=0.4\linewidth]{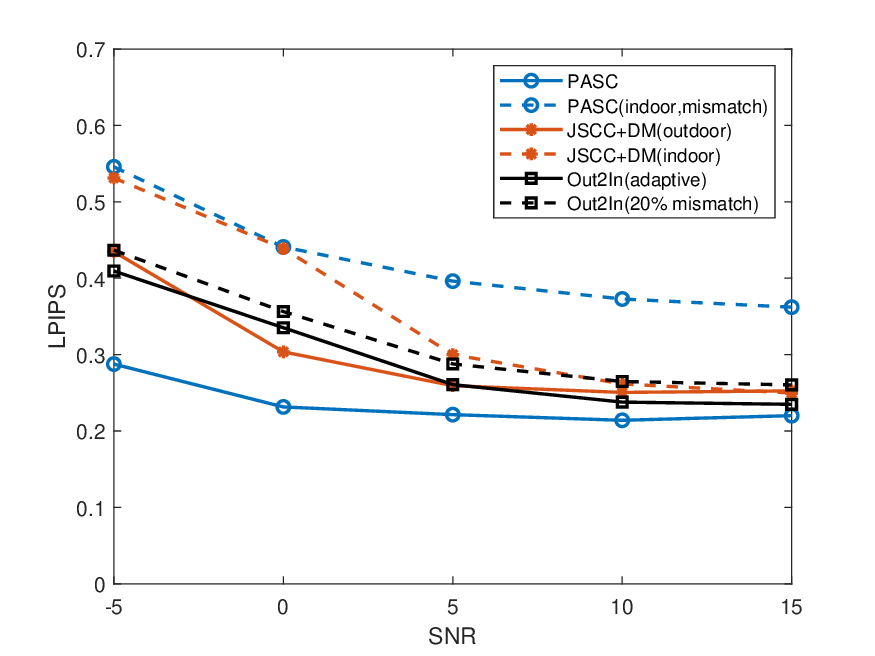}}}
	\caption{(a) The ratio of zero pixels between matched and mismatched images. More zero pixels under the specific threshold save bandwidth. (b) LPIPS performance of the competing methods.}
	\label{SecC_per}
\end{figure*}
\begin{figure*} 
	\centering

	\subfigure[Target image]{
		\includegraphics[width=0.22\linewidth]{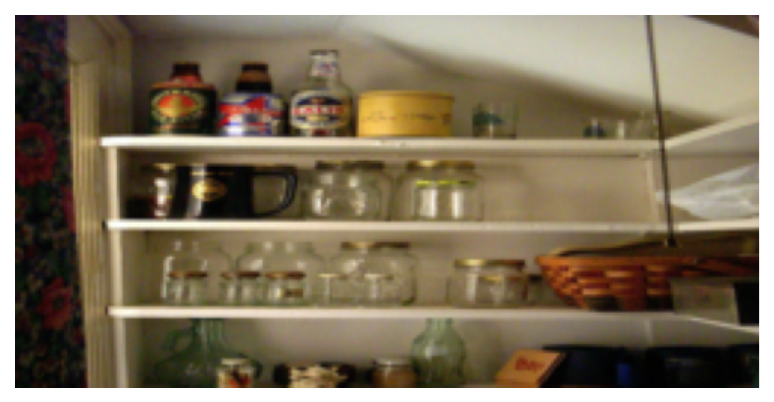}}
	\subfigure[JSCC, SNR=10 dB]{
		\includegraphics[width=0.22\linewidth]{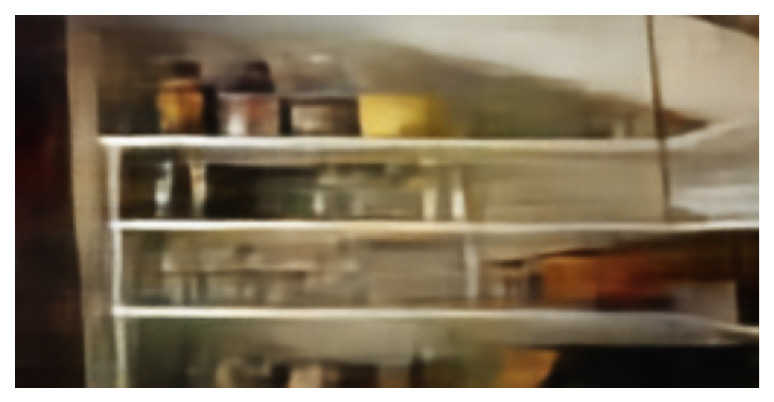}}	
	\subfigure[JSCC+DM, SNR=10 dB]{
		\includegraphics[width=0.22\linewidth]{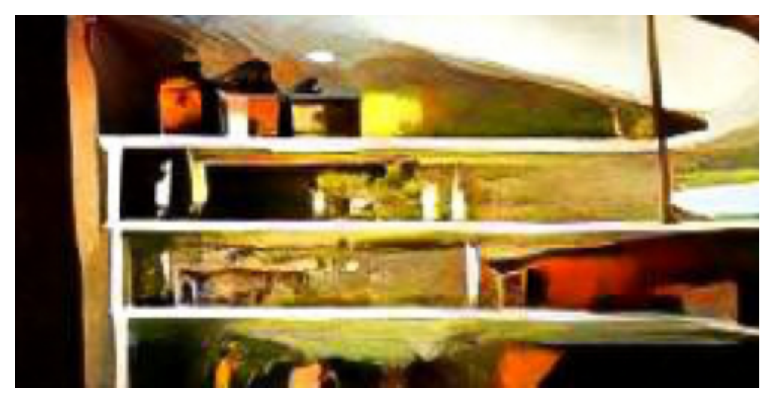}}
	\subfigure[PASC(mismatch), SNR=10 dB]{
		\includegraphics[width=0.22\linewidth]{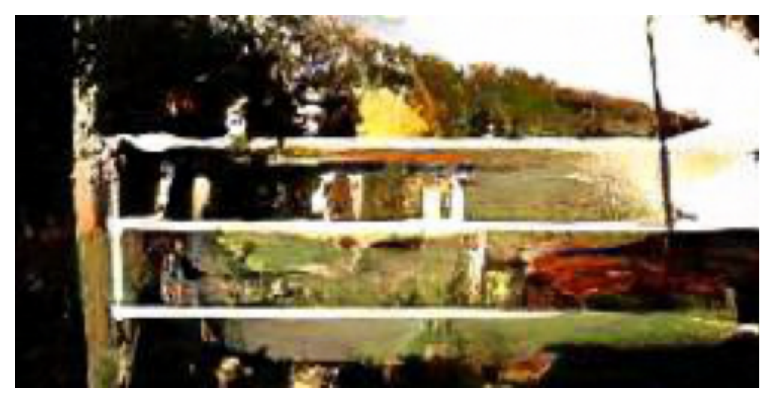}}

	\caption{Examples of different methods when transmitting indoor images using the outdoor strategy, where PASC is disturbed by this mismatch. In this case, the JSCC methods are robust to this change.}
	\label{EX_C}
\end{figure*}

To illustrate the effectiveness of the LLM recommendations, we input the simulation results from Figs. \ref{SecA_per} and \ref{SecB_per} as a performance list. The LLM is then instructed to select the best method to meet the required performance. For example, we select a scenario where LPIPS performance should be better than 0.23. The first example in TABLE \ref{Metric1} directs the LLM to choose a method with the lowest complexity. The results show that the LLM uses 16k bits and selects the JSCC method to minimize complexity when SNR $>$ 5 dB. However, when SNR $<$ 0 dB, JSCC cannot meet the performance requirement, making PASC the optimal choice.
The second example prioritizes reducing bandwidth consumption. Thus, PASC is consistently chosen when SNR $\leq$ 0 dB. At 5 dB SNR, JSCC(16k) is optimal, as PASC(16k) requires the same bandwidth but has higher complexity. At 10 dB SNR, PASC(2k) reduces bandwidth while still meeting performance requirements.
In the third example, the LLM suggests settings that are not present in the current performance data. Unlike the second example, the LLM recommends using $\epsilon=1$ and 8k bits for PASC at $-5$ dB SNR to improve performance. It also suggests using 8k bits for PASC at $5$ dB SNR to optimize bandwidth efficiency. These settings were used to train and apply new encoder-decoders. The results show improved performance at $-5$ dB SNR, although still below 0.23. At 5 dB SNR, bandwidth is saved, albeit with slightly reduced performance.

This suggests that the performance list may be insufficient, as it has only been tested with 16k, 2k, and 1k bandwidths. In a practical system, the recommended settings can be tested using images known to both the transmitter and receiver, once the corresponding encoder-decoders are trained. These performances can then be added to the list to improve future LLM recommendations. The above experiments demonstrate the superiority of the proposed method in improving task performance and resource utilization.

\subsection{Performances Under Mismatch Conditions}
In this subsection, we first introduce how to detect mismatches caused by outdated or incorrect KBs. We then compare the performance of competing methods under mismatched conditions.

\begin{figure*} 
	\centering
	\includegraphics[width=0.9\linewidth]{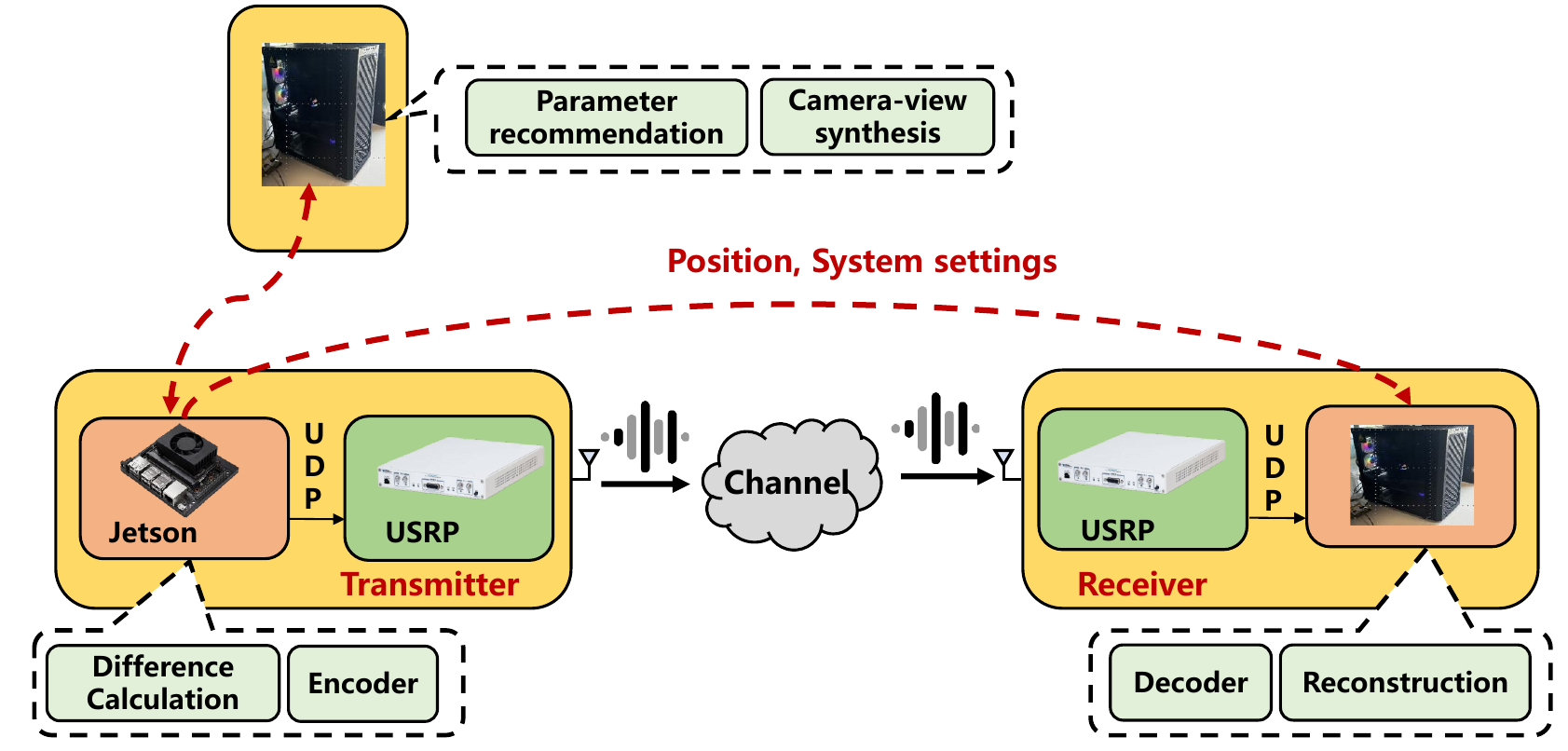}
	\caption{Hardware implementation structure of the proposed method. The Jetson+USRP acts as a compact device at the server, while the server+USRP functions as a BS.}
	\label{hardware}
\end{figure*}

Fig. \ref{SecC_per}(a) shows the ratio of zero pixels when the threshold $\epsilon = 0.4$, which indicates the amount of transmission bandwidth saved by omitting zero-valued pixels. In an ``Outdoor Match'' scenario, where positional information is accurate, PASC methods perform optimally. The synthetic image, based on position, closely resembles the transmitted image, resulting in many pixels in the transmitted image being set to zero. Conversely, in the ``Outdoor Mismatch'' scenario, where positional information is outdated, the synthetic and transmitted images do not match. However, about half the pixels can still be ignored due to shared similarities, such as sky and road elements, between the transmitted and synthetic images. In the ``Indoor Mismatch'' scenario, indoor images are not position-related, and very few pixels match the synthetic image. When transitioning from outdoor to indoor images without timely updated information, performance is affected. Since FMs are computationally complex and suffer from transmission delays, it is not always possible to access them in time. Therefore, we propose a local strategy to guarantee transmission performance by applying PASC only when the ratio of zero pixels exceeds 0.45, effectively avoiding most mismatches without relying on FMs.

This strategy is implemented in Fig. \ref{SecC_per}(b) as ``Out2In (adaptive),'' where JSCC+DM is used for indoor images, and PASC is used for outdoor images. Due to FM decision-making delays, some indoor images are still transmitted using PASC, resulting in slightly inferior performance, denoted as ``Out2In (mismatch),'' compared to ``Out2In (adaptive).'' PASC achieves optimal performance for outdoor images with accurate position information. However, for indoor images, where position information is irrelevant, PASC(indoor, mismatch) performs poorly. JSCC+DM transmits the entire image, effectively restoring both indoor and outdoor images at high SNRs. However, when the SNR is low, the performance gap between JSCC+DM (indoor) and JSCC+DM (outdoor) widens, as indoor images are more complex and harder to transmit under poor channel conditions.

The examples in Fig. \ref{EX_C} highlight the effects of mismatches. The indoor image contains various objects, such as bottles and glasses, making it complex. The image processed by JSCC appears blurry, but DM further refines it, albeit with some introduced changes. However, when the DM is disturbed by irrelevant positional information, the reconstruction quality is significantly reduced.

The experiments demonstrate that the proposed method effectively leverages position information to enhance transmission performance. Moreover, FM-based reconstruction and method recommendation can adaptively switch between methods to meet transmission requirements under varying image content and channel conditions. Finally, to address mismatches caused by outdated or incorrect information from FMs, a simple local judgment can be applied to ensure reasonable performance.

\section{Testbed Implementation and Discussion}
Prototype verification of semantic communications is a crucial step toward achieving real-time transmission. In this section, we first describe the implementation details, focusing on establishing a proper transmission environment with limited bandwidth and computational resources. Next, we discuss the OTA results and analyze the challenges of real-time transmission.

\subsection{Implementation Details}
As shown in Fig. \ref{hardware}, the prototype developed from our earlier work \cite{ding2024adaptive} consists of one NVIDIA Jetson Xavier NX, two USRP-2943R devices (each equipped with one antenna), and a server. Unlike previous work, the prototype has been reassembled to support FM-based methods. The transmission process simulates a camera sending images to a BS. The camera terminal is represented by the low-power Jetson device, while the BS is represented by a powerful server. The server, powered by an Intel i9-14900k CPU and an Nvidia RTX 4090 GPU, can handle FM operations. Although the Jetson Xavier NX, with its 21 TOPS AI performance, has a GPU, it struggles to run large models, so compact AI modules (e.g., the encoder) are processed locally, while the FM-based tasks are offloaded to the external server.

Both the server and Jetson Xavier NX are programmed in Python and Pytorch, facilitating easy deployment of AI modules. At the transmitter, the Jetson Xavier NX retrieves the shared KB from the server, performs the difference calculation, and runs the encoder. The codeword is sent from the Jetson to the USRP using user datagram protocol (UDP). The USRP handles signal processing at the physical layer, using a OFDM setup as in \cite{ding2024adaptive}. Each OFDM frame has a length of 13 ms, and the effective throughput is approximately 1.4 Mbps. After the bitstream is recovered by the second USRP, it is sent to the server via UDP, where the shared KB and semantic reconstruction are applied to restore the image.

\begin{figure} 
	\centering
	\includegraphics[width=0.90\linewidth]{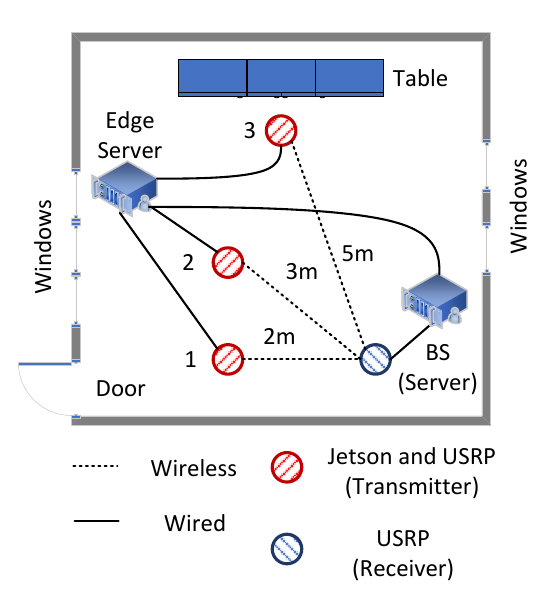}
	\caption{OTA scenarios of the proposed prototype.}
	\label{scen}
\end{figure}

\subsection{OTA Results}

Using the aforementioned hardware, the OTA scenarios were established as shown in Fig. \ref{scen}. Capturing real-world images with their positions is challenging, so the transmitted images and positional coordinates were randomly selected from existing datasets. The transmitter was then moved to test performance under different channel conditions. Communication between the Jetson, server, and USRP relies on wired connections. The transmitter regularly shares its position and system settings with the receiver, with negligible overhead compared to image transmission. However, transmission delays are significant because matched KBs are essential for the proposed semantic methods, as discussed in Section III.D. If the Jetson directly queries the LLM for parameter recommendations or calls the DM for view synthesis as part of the shared KB, the process can take several seconds, making frequent updates impractical. Instead, using a compact UNet for local view synthesis is a better solution, as it is faster and only requires position coordinates.

 \begin{table} 
 	\centering	
 	 \footnotesize
 	 \caption{Runtime of different modules on different devices.}
 	
 	\begin{tabular}{lll}
 		\toprule
 		Module&Jetson &Server \\\midrule
 		Encoder& 7.7 ms & 1.3 ms \\ \midrule
 		Decoder& 1.2 ms & <1 ms \\ \midrule
 		UNet (view-synthesis) & 106 ms& 4.4 ms\\ \midrule
 		DM (view-synthesis/ & /& 3.4 s\\
 		semantic reconstruction) &&\\ \midrule
 		LLM (scenario detection  & /& 2.1 s\\
 		/parameter recommendation)&&\\ \midrule
 		Scenario detection (local) &4 ms &<1ms\\ 	
 		\bottomrule \\
 		\multicolumn{3}{l}{*Server is equipped with a CPU (Intel i9-14900k) and}\\
 		\multicolumn{3}{l}{~~a GPU (Nvidia RTX 4090).} \\
 		\multicolumn{3}{l}{*Jetson Xavier NX has a good performance (21 TOPS)}\\
 		\multicolumn{3}{l}{~~with low power consumption (about 10 watts). } \\
 		\multicolumn{3}{l}{*Sharing position coordinate from transmitter to receiver}\\
 		\multicolumn{3}{l}{~~needs about 5 ms in our prototype. }
 	\end{tabular}
 	\label{Metric2}
 \end{table}

Table \ref{Metric2} shows the runtime of different modules. The most complex modules, such as the LLM and DM, are unsuitable for real-time use and are therefore executed every 5 seconds to provide optimal parameter recommendations. The position coordinates are updated at the transmitter and shared with the receiver, with a transmission delay of about 5 ms. However, even the UNet-based view synthesis, while faster than DM, requires about 106 ms to generate a camera-view image, so the same camera-view image is reused during this time window. This is acceptable, as position changes within such a short time are minimal. The total runtime of other modules on the Jetson is less than 10 ms. On the receiver side, the decoder takes less than 1 ms, but DM-based reconstruction remains a challenge. Despite reducing the number of diffusion steps to 50, it still takes approximately 3 seconds to process an image. This time cost is a major constraint on applying the PASC method. However, acceleration techniques like distillation \cite{meng2023distillation,song2023consistency} could further reduce diffusion steps. Given that OFDM transmission can support hundreds of JPEG+LDPC-encoded images or around 40 PASC-encoded images per second, the advantages of PASC are still significant.

  \begin{table} 
 	\centering	
    \footnotesize
 	\caption{LPIPS performance of the proposed method in our prototype.}
 	
 	\begin{tabular}{clll}
 		\toprule
 		LPIPS&Proposed &JSCC &JPEG+LDPC \\\midrule
 		High SNR& \textbf{0.183 }& 0.183 &0 \\ \midrule
 		Moderate SNR& \textbf{0.254} & 0.318 &0.873\\ \midrule
 		Low SNR & \textbf{0.329}&0.561 &1.13\\
 		\bottomrule
 		
 	\end{tabular}
 	\label{Metric3}
 \end{table}

To further evaluate the methods, we tested the prototype in three scenarios (Fig. \ref{scen}), adjusting the distance between the transmitter and receiver to simulate low, moderate, and high SNRs. The system settings were automatically adjusted within a few seconds as the transmitter moved. As shown in Table \ref{Metric3}, the proposed method consistently has a better LPIPS performance than the competing methods. In the high SNR scenario, JSCC was selected due to sufficient bandwidth, conserving computational resources while yielding optimal performance for both JSCC and the proposed method. In moderate and low SNR scenarios, the PASC method was used, achieving better performance than JSCC. The conventional JPEG+LDPC method failed to transmit images effectively under these conditions.

Overall, the proposed method proves effective in our prototype, with real-time constraints taken into account. FM-based semantic communication using positional information will become even more competitive as FM acceleration techniques advance.

\section{Conclusion}
\label{s6}

This study explores the use of positional information for semantic image transmission, leveraging the strong relationship between outdoor images and their corresponding positions. The proposed framework, called PASC, effectively utilizes state-of-the-art FMs to convert positional data into synthetic images, enhancing the transmission process. Key components of the framework---view synthesis, semantic encoder-decoder, and DM-based reconstruction---are developed to ensure efficient image transmission, even under limited bandwidth and low SNR conditions.
An LLM-based adaptive strategy is also introduced, allowing for system parameter recommendations, including network selection, code length, and mask threshold, based on changing scenarios and requirements. Furthermore, real-time implementation challenges are addressed, with attention to potential delays that may cause outdated KBs and errors introduced by FMs, which can affect performance. To ensure robustness against these mismatches or errors, low-complexity local networks are designed.
Finally, the runtime of the various modules was tested on our hardware prototype, demonstrating that the proposed framework performs well across different scenarios, validating its effectiveness in practical, real-time applications.

	\bibliographystyle{IEEEtran}
	\bibliography{bibtex0320}
	
	%
	
	
	
\end{document}